\newcount\mgnf\newcount\tipi\newcount\tipoformule
\newcount\aux\newcount\piepagina\newcount\xdata
\mgnf=0
\aux=0           
\tipoformule=1   
\piepagina=1     
\xdata=1         
\def\Di{27 luglio 2000}

\ifnum\mgnf=1 \aux=0 \tipoformule =1 \piepagina=1 \xdata=1\fi

\def\9#1{\ifnum\aux=1#1\else\relax\fi}
\ifnum\piepagina=0 \footline={\rlap{\hbox{\copy200}\
$\st[\number\pageno]$}\hss\tenrm \foglio\hss}\fi \ifnum\piepagina=1
\footline={\rlap{\hbox{\copy200}} \hss\tenrm \folio\hss}\fi
\ifnum\piepagina=2\footline{\hss\tenrm\folio\hss}\fi

\ifnum\mgnf=0 \magnification=\magstep0
\hsize=12.5truecm\vsize=23truecm \parindent=4.pt\fi
\ifnum\mgnf=1 \magnification=\magstep1
\hsize=16.0truecm\vsize=22.5truecm\baselineskip16pt\vglue5.0truecm
\overfullrule=0pt \parindent=4.pt\fi

\let\a=\alpha\let\b=\beta  \let\d=\delta
\let\e=\varepsilon \let\z=\zeta \let\h=\eta
\let\k=\kappa \let\l=\lambda \let\m=\mu \let\n=\nu
\let\x=\xi \let\p=\pi \let\r=\rho \let\s=\sigma 
 \let\f=\varphi \let\ps=\psi \let\o=\omega
 \let\G=\Gamma \let\D=\Delta 
\let\L=\Lambda   \let\F=\Phi
  
{\count255=\time\divide\count255 by 60 \xdef\oramin{\number\count255}
\multiply\count255 by-60\advance\count255 by\time
\xdef\oramin{\oramin:\ifnum\count255<10 0\fi\the\count255}}
\def\ora{\oramin }

\ifnum\xdata=0
\def\data{\number\day/\ifcase\month\or gennaio \or
febbraio \or marzo \or aprile \or maggio \or giugno \or luglio \or
agosto \or settembre \or ottobre \or novembre \or dicembre
\fi/\number\year;\ \ora}
\def\Di{\number\day\kern2mm\ifcase\month\or gennaio \or
febbraio \or marzo \or aprile \or maggio \or giugno \or luglio \or
agosto \or settembre \or ottobre \or novembre \or dicembre
\fi\kern0.1mm\number\year}
\else
\def\data{\Di}
\fi

\setbox200\hbox{$\scriptscriptstyle \data $}
\newcount\pgn \pgn=1
\def\foglio{\number\numsec:\number\pgn
\global\advance\pgn by 1} \def\foglioa{A\number\numsec:\number\pgn
\global\advance\pgn by 1}
\global\newcount\numsec\global\newcount\numfor \global\newcount\numfig
\gdef\profonditastruttura{\dp\strutbox}
\def\senondefinito#1{\expandafter\ifx\csname#1\endcsname\relax}
\def\SIA #1,#2,#3 {\senondefinito{#1#2} \expandafter\xdef\csname
#1#2\endcsname{#3} \else \write16{???? ma #1,#2 e' gia' stato definito
!!!!} \fi} \def\etichetta(#1){(\veroparagrafo.\veraformula) \SIA
e,#1,(\veroparagrafo.\veraformula) \global\advance\numfor by 1
\9{\write15{\string\FU (#1){\equ(#1)}}} \9{ \write16{ EQ \equ(#1) == #1
}}} \def \FU(#1)#2{\SIA fu,#1,#2 }
\def\etichettaa(#1){(A\veroparagrafo.\veraformula) \SIA
e,#1,(A\veroparagrafo.\veraformula) \global\advance\numfor by 1
\9{\write15{\string\FU (#1){\equ(#1)}}} \9{ \write16{ EQ \equ(#1) == #1
}}} \def\getichetta(#1){Fig.  \verafigura \SIA e,#1,{\verafigura}
\global\advance\numfig by 1 \9{\write15{\string\FU (#1){\equ(#1)}}} \9{
\write16{ Fig.  \equ(#1) ha simbolo #1 }}} \newdimen\gwidth \def\BOZZA{
\def\alato(##1){ {\vtop to \profonditastruttura{\baselineskip
\profonditastruttura\vss
\rlap{\kern-\hsize\kern-1.2truecm{$\scriptstyle##1$}}}}}
\def\galato(##1){ \gwidth=\hsize \divide\gwidth by 2 {\vtop to
\profonditastruttura{\baselineskip \profonditastruttura\vss
\rlap{\kern-\gwidth\kern-1.2truecm{$\scriptstyle##1$}}}}} }
\def\alato(#1){} \def\galato(#1){}
\def\veroparagrafo{\number\numsec}\def\veraformula{\number\numfor}
\def\verafigura{\number\numfig}
\def\geq(#1){\getichetta(#1)\galato(#1)}
\def\Eq(#1){\eqno{\etichetta(#1)\alato(#1)}}
\def\eq(#1){\etichetta(#1)\alato(#1)}
\def\Eqa(#1){\eqno{\etichettaa(#1)\alato(#1)}}
\def\eqa(#1){\etichettaa(#1)\alato(#1)}
\def\eqv(#1){\senondefinito{fu#1}$\clubsuit$#1\write16{No translation
for #1} \else\csname fu#1\endcsname\fi}
\def\equ(#1){\senondefinito{e#1}\eqv(#1)\else\csname e#1\endcsname\fi}
\openin13=#1.aux \ifeof13 \relax \else \input #1.aux \closein13\fi
\openin14=\jobname.aux \ifeof14 \relax \else \input \jobname.aux
\closein14 \fi \9{\openout15=\jobname.aux} \newskip\ttglue

\font\titolo=cmbx10 scaled \magstep1
\font\ottorm=cmr8\font\ottoi=cmmi7\font\ottosy=cmsy7
\font\ottobf=cmbx7\font\ottott=cmtt8\font\ottosl=cmsl8\font\ottoit=cmti7
\font\sixrm=cmr6\font\sixbf=cmbx7\font\sixi=cmmi7\font\sixsy=cmsy7

\font\fiverm=cmr5\font\fivesy=cmsy5\font\fivei=cmmi5\font\fivebf=cmbx5
\def\ottopunti{\def\rm{\fam0\ottorm}\textfont0=\ottorm%
\scriptfont0=\sixrm\scriptscriptfont0=\fiverm\textfont1=\ottoi%
\scriptfont1=\sixi\scriptscriptfont1=\fivei\textfont2=\ottosy%
\scriptfont2=\sixsy\scriptscriptfont2=\fivesy\textfont3=\tenex%
\scriptfont3=\tenex\scriptscriptfont3=\tenex\textfont\itfam=\ottoit%
\def\it{\fam\itfam\ottoit}\textfont\slfam=\ottosl%
\def\sl{\fam\slfam\ottosl}\textfont\ttfam=\ottott%
\def\tt{\fam\ttfam\ottott}\textfont\bffam=\ottobf%
\scriptfont\bffam=\sixbf\scriptscriptfont\bffam=\fivebf%
\def\bf{\fam\bffam\ottobf}\tt\ttglue=.5em plus.25em minus.15em%
\setbox\strutbox=\hbox{\vrule height7pt depth2pt width0pt}%
\normalbaselineskip=9pt\let\sc=\sixrm\normalbaselines\rm}

\catcode`@=11
\def\footnote#1{\edef\@sf{\spacefactor\the\spacefactor}#1\@sf
\insert\footins\bgroup\ottopunti\interlinepenalty100\let\par=\endgraf
\leftskip=0pt \rightskip=0pt \splittopskip=10pt plus 1pt minus 1pt
\floatingpenalty=20000
\smallskip\item{#1}\bgroup\strut\aftergroup\@foot\let\next}
\skip\footins=12pt plus 2pt minus 4pt\dimen\footins=30pc\catcode`@=12
\let\nota=\ottopunti

\newdimen\xshift \newdimen\xwidth \newdimen\yshift

\def\ins#1#2#3{\vbox to0pt{\kern-#2 \hbox{\kern#1
#3}\vss}\nointerlineskip}

\def\eqfig#1#2#3#4#5{ \par\xwidth=#1
\xshift=\hsize \advance\xshift by-\xwidth \divide\xshift by 2
\yshift=#2 \divide\yshift by 2 \line{\hglue\xshift \vbox to #2{\vfil #3
\includegraphics{#4.ps} }\hfill\raise\yshift\hbox{#5}}} 

\def\dsgn#1#2#3#4{\xshift=#1%
\yshift=#2 \divide\yshift by 2 \hbox{ \vbox to #2{\vfil #3%
\includegraphics{#4.ps} }}\kern\xshift}

\def\8{\write13}


\def\V#1{{\,\underline#1\,}}
\def\T#1{#1\kern-4pt\lower9pt\hbox{$\widetilde{}$}\kern4pt{}}
\let\dpr=\partial\def\Dpr{{\V\dpr}} \let\io=\infty\let\ig=\int
\def\fra#1#2{{#1\over#2}}\def\media#1{\langle{#1}\rangle}\let\0=\noindent
\def\guida{\leaders\hbox to 1em{\hss.\hss}\hfill}
\def\tende#1{\vtop{\ialign{##\crcr\rightarrowfill\crcr
\noalign{\kern-1pt\nointerlineskip} \hglue3.pt${\scriptstyle
#1}$\hglue3.pt\crcr}}} \def\otto{\
{\kern-1.truept\leftarrow\kern-5.truept\to\kern-1.truept}\ }

\def\pagina{\vfill\eject}

\def\st{\scriptscriptstyle}
\def\*{\vskip0.3truecm}

\def\lis#1{{\overline #1}}\def\eg{\hbox{\it e.g.\ }}

\def\ie{\hbox{\it i.e.\ }}

\def\fiat{{}}
\def\\{\hfill\break} \def\={{ \; \equiv \; }}

\ifnum\aux=1\BOZZA\else\relax\fi
\ifnum\tipoformule=1\let\Eq=\eqno\def\eq{}\let\Eqa=\eqno\def\eqa{}
\def\equ{{}}\fi
\def\defi{\,{\buildrel def \over =}\,}

\def\1{\ifnum\mgnf=0\pagina\else\relax\fi}
\def\W#1{#1_{\kern-3pt\lower6.6truept\hbox to 1.1truemm
{$\widetilde{}$\hfill}}\kern2pt\,}

\def\FINE{
\*
\0{\it
Preprints at: {\tt http://ipparco.roma1.infn.it}\\
\sl e-mail: {\tt gallavotti@roma1.infn.it}
}}

\def\xx{{\V x}}

\def\LL{{\cal L}}

\def\oo{{\V \o}}\def\kk{{\V k}}

\def\ndpr{{\kern1pt\raise 1pt\hbox{$\not$}\kern1pt\dpr\kern1pt}}
\def\Ndpr{{\kern1pt\raise 1pt\hbox{$\not$}\kern0.3pt\dpr\kern1pt}}

\def\cfr{{\it cf.\ }}
\fiat

\fiat
\def\pst{{\tilde\ps}}\def\Tr{{\,\rm Tr\,}}

\0{\titolo The Luttinger model: its role in the RG--theory}\\
{\titolo of one dimensional many body Fermi systems}
\*
\0{\bf Giovanni Gallavotti}\\
\0{\it Fisica, Universit\`a di Roma}\\
\0{\it P.le A. Moro 2, 00185 Roma, Italia}\\
\0{e-mail: \tt gallavotti@roma1.infn.it}
\*
\*

\0{\bf Abstract:} {\it The Luttinger model was introduced to illustrate
the theory of Tomonaga via an exactly soluble model. It became soon
the subject of great interest also on the part of Mathematical Physics
and a key to the investigations of the mathematical properties of
Condensed Matter Physics. This paper reviews aspects of the above
developments relevant for renormalization group methods.}

\*
\0{\bf\S1. The Luttinger model.}
\numsec=1\numfor=1
\*

The model describes many body systems of two kinds of fermions on a
line whose fields are $\pst^\a_{\o,x}$, with $\a=\pm$ specifying the
creation and annihilation operators for fermions located at a point
$x\in [-L/2,L/2]\subset R^1$ and distinguished by the label $\o=+$ or
$\o=-$. The Hamiltonian, ``(kinetic energy)+ (chemical
potential)+(potential energy)'', is written as

$$\eqalign{
&H=\ig_{-L/2}^{L/2} dx\, \sum_{\o=\pm}
\pst^+_{\o,x}v_F (i\o\dpr_x-p_F)\pst^-_{\o,x}+\cr
&+\l \ig_{-L/2}^{L/2} dx dy\,
\pst^+_{+,x}\pst^-_{+,x}\pst^+_{-,y}\pst^-_{-,y}\, v(x-y)\cr}\Eq(1.1)$$
where $\l$ is a coupling constant and $v(x-y)$ is a smooth short range
pair potential (\eg $v(x-y)\=0$ if $|x-y|\ge p_0^{-1}$ for some
``range'' $p_0^{-1}$) . The parameter $v_F$ is the ``velocity at the
Fermi surface'', $v_F=p_F/m$ if $m$ is the mass of the particles: we
have set $\hbar=m=1$.

This model was introduced in [Lu63] and it illustrated Tomonaga's
theory of spin $0$ fermions in one dimension, [To50], which shows the
remarkable phenomenon of the {\it anomaly} of the ground state: \ie a
ground state with a density of states which does not have a
discontinuity at the Fermi momentum $k=p_F$ but its graph becomes
vertical with a vertical tangency exponent $a(\l)=O(\l^2)$ called the
{\it anomaly}'' of the Fermi surface.
\*

The model supposes {\it a priori} that the ``physically significant''
part of the Hamiltonian is described by {\it quasi particles}.
This means realizing that the Schwinger functions of the ground state
of a spin $0$ Fermi gas can be identified, to leading non trivial order
in the coupling constant $\l$, by thinking that the system in fact
consists of two particles with energy close to the Fermi energy
$p_F^2/2$, one with momentum close to $+p_F$ and the other with
momentum close to $-p_F$ whose fields, in a suitable superposition,
yield the field $\pst^\pm_x$ of the observable particles.

Writing a momentum close to $\pm p_F$ as $\pm p_F+ k$
the free field $\pst^\pm_x$ is expressed as

$$\pst^\pm_{x}\simeq \sum_{\o=\pm} \ps^\pm_{\o,x} e^{\pm i\,p_F\,\o
x}\Eq(1.2)$$
where calling $a^{\pm}_{k,\o} \defi \a^{\pm}_{k+p_F\o,\o}$, if
$\a^\pm_p$ are the creation and annihilation operators for the Fermi
particles, the {\it quasi particles fields} are

$$\ps^{\pm}_{\o,x}=
{1\over \sqrt{L}} \sum_{k} e^{\pm ikx} a^{\pm}_{\o,k}\Eq(1.3)$$
Hence if the operator $T_0$

$$T_0 \defi \sum_\o\sum_{ k >0}  v_F\,k\,  (a^+_{\o k ,\o} a^-_{\o k ,\o} +
a^-_{-\o k ,\o} a^+_{-\o k ,\o} )\Eq(1.4)$$
approximates the free fermionic kinetic energy plus the chemical
potential energy for $k\simeq 0$,  then the quasi particle fields at time $t$
will be $\ps^\pm_{\o,x,it}=e^{\pm i T_0 t}\ps^\pm_{\o,x}e^{\mp i T_0
t}$ $=e^{\pm i p_F t}\ps_{\o,x,it}^\pm$ with

$$\ps^{\pm}_{\o,x,t} \equiv e^{tT_0}\ps^{\pm}_{x,\o}e^{-tT_0} =
{1\over \sqrt{L}} \sum_{ k} e^{\pm (i\, k\, x +\,t\,\o\, k)} a^{\pm}_{ k,\o}
\Eq(1.5)$$

I do not repeat here the heuristic analysis showing that taking the
two kinds of particles Hamiltonian \equ(1.1) and approximating it with
the free kinetic energy \equ(1.4) and expressing the fermionic field
$\pst^\pm$ via \equ(1.2) means
\\ (1) thinking the particles with
positive momentum as distinct from the ones with negative momentum
and, {\it furthermore},
\\ (2) replacing the dispersion relation
$\e(p)=p^2/2$ with $p=\pm p_F+k$ with $\e(\pm,k)= p_F^2/2\pm k p_F$
and, also,
\\ (3) allowing $k$ to take all values rather than $k>-p_F$
or $k<p_F$.
\*

Luttinger realized that Tomonaga's theory amounted, to leading order
in $\l$, to

\0(i) replacing the real fermionic particle fields with the
$\pst^\pm$ fields defined by th r.h.s. of \equ(1.2) and
\equ(1.3),

\0(ii) replacing the ``real'' fermionic particle kinetic energy
defined as

$$T'_0=\ig_{-L/2}^{L/2} dx\, \pst^+_x \,\fra12( \dpr_x^2-
p_F^2)\pst^-_x\=\sum_p \fra12(p^2-p_F^2) \,\a^+_p\a^-_p \Eq(1.6)$$
where $p=2\p n/L$, $n=0,\pm1,\pm2,\ldots$, by the operator $T_0$ above,
which can be written

$$T_0=\ig_{-L/2}^{L/2} dx\, \sum_\o
\ps^+_{\o,x}(-iv_F\o\dpr_x)\ps^-_{\o,x}
\Eq(1.7)$$
{\it and}

\0(iii) replacing the potential energy by the expression in \equ(1.1), {\it
rather than} considering the usual pair potential which would be
written as
$$\l\ig_{-L/2}^{L/2}
dx\,dy\,\pst^+_x\pst^-_x\,\pst^+_y\pst^-_y\,v(x-y)\Eq(1.8)$$
Note that comparing \equ(1.8) with the potential energy in \equ(1.1)
and with \equ(1.2) several ``cross terms'' involving fields with
different quasi particles labels are absent.
\*

The Luttinger approximation was stated to be exactly soluble in the
remarkable paper [Lu63] and the solution did yield the expected
anomaly of the Fermi surface mentioned above, thereby providing a
simple explanation of the phenomenon.

The exact solution was, however, not really correct because of an
error on the Fock space canonical commutation relations which, in
infinitely many degrees of freedom systems, do not have a unique
representation: a matter that is now well understood but that was not
so clear at the time. Nevertheless the attempt in [Lu63] contained the
important ideas indicating strongly that the model could probably be
really solved exactly. The exact solution was discovered a little later by
Mattis and Lieb, [ML65]. The real value of the anomaly and Luttinger's
agree to leading order in $\l$ (\ie to second order) but they differ
in the higher orders (which vanish identically in Luttinger's
expression while in [ML65] the anomaly is analytic in $\l$ near $\l=0$
but {\it it is not} a polynomial of second order); see also [BGM92].

The work [ML65] should be seen as a part of the series of exact
solutions of simple models which, starting with the $2$--dimensional
Ising model led to the Bethe ansatz (which is not necessary in the
case of the Luttinger model) and to exactly solvable $6$ and $8$
vertex models, [ML66], [Ba82].

The Luttinger model is closely related to the Thirring model of
quantum field theory, [Lu63]: for an analysis of the relation between
the two models see [Ma94].
\*

\0{\bf\S2. Schwinger functions and ground state.}
\numsec=2\numfor=1
\*

More recently there has been renewed interest, particularly after the
discovery of high temperature superconductivity, in understanding
the properties of condensed matter from a fundamental point of view,
\ie without relying on heuristic arguments, whenever possible.
One of the directions in which research has been stimulated is the
application of multiscale analysis, \ie of the renormalization group
methods, to investigate the large distance properties of the
correlations in the ground states of the simplest systems, \ie Bose or
Fermi gases with weak short range interactions: [BG90], [FT90],
[BGPS94], [Sh92], [BG95]. Other methods have also been applied leading
to complementary results valid also at strong coupling, for instance
[KL73] and [KLY88], [LY00], or for exact solutions [LL63], [LW68].
\*

A mathematically rigorous theory of the ground state of the
$1$--dimensional Fermi gas turned out surprisingly hard even at small
$\l$, probably because the basic formalism, even today, is still
waiting to acquire an established shape. Furthermore the case of
spinning electrons with attractive rotationally symmetric pair
interaction still defies research. The situation is worse in higher
dimension where only formal developments seem to be available (see
[BG90], [FT90], [Sh92]): therefore higher dimension will be left out
of the present discussion, but for sporadic comments.

In dimension $1$ the Hamiltonian of a system of spin $0$, mass $m=1$,
fermionic particles in a periodic box $[-L/2,L/2]$ (and in the grand
canonical ensemble) is

$$ H=\sum_{s=1}^N\left(-\fra12\Delta_{x_i}-\m\right) + 2
\l\sum_{i<j} v(x_i-x_j) \Eq(2.1)$$
with $N$=number of particles or equivalently, in second quantized form,

$$\eqalign{
H=&\ig_{-L/2}^{L/2} dx\, \pst^+_x(-\fra12\D_x-\m)\pst^-_x+\cr
&+ \l \ig_{-L/2}^{L/2} dx\, dy\, v(x-y)
\,\pst^+_x\pst^-_x\,\pst^+_y\pst^-_y
\cr}\Eq(2.2)$$
where $\pst$ is the usual Fermi field (rather than its approximation
in \equ(1.2), \equ(1.3)).

If $\r$ is the density of the gas the simplest question that one can
ask is about the behavior near $p=\pm \p \r\defi\pm p_F=\pm\sqrt{2\m}$ of the
Fourier transform of the one particle reduced density matrix.  Note
that if $\l=0$ it is known that the one particle reduced density
matrix at inverse temperature $\b=+\io$ is

$$\media{\pst^+_x\pst^-_0}\defi
\lim_{\b\to\io} \fra{\Tr e^{-\b H}\pst^+_x\pst^-_0}{\Tr e^{-\b H}}\Eq(2.3)$$
with Fourier transform $\chi_{p_F}(p)=1$ if $|p|<p_F$ and zero otherwise.

In the following it will be more convenient to study the more general
Schwinger functions

$$\eqalignno{
&S_{\s_1\cdots\s_n}(t_1x_1,\ldots,t_n{x_n})=(-1)^\p
\lim_{\b\to\io}\lim_{\L\to\io} \cdot &\eq(2.4)\cr
\cdot&{\Tr \left( e^{-(\b-t_{\p(1)})H} \psi^{\s_{\p(1)}}_{x _{\p(1)}}
e^{-(t_{\p(1)}-t_{\p(2)})H} \psi^{\s_{\p(2)}}_{x _{\p(2)}}\cdots
\psi^{\s_{\p(n)}}_{x _{\p(n)}}e^{- t_{\p(n)} H}\right)\over \Tr\, e^{-\b H}}
\ ,\cr}$$
where $\s_i=\pm1$, $\p$ is the permutation of $(1,\ldots,n)$, such
that $t_{\p(1)} > t_{\p(2)} > \ldots > t_{\p(n)}$ and $(-1)^\p$ is the
permutation parity; $n$ is even. In this way $\chi_{p_F}(p)$ is the
Fourier transform of $S_{-+}((0^-,x),(0,0))$.
\*

\0{\bf\S3. Functional integration for $1$--dim spinless
fermions. Luttinger--Kohn--Ward determination of the chemical potential.}
\numsec=3\numfor=1\*

Denoting $\x=(x_0,x), \h=(y_0,y)\in R^2,\k=(k_0,k)\in R^2$ the
Schwinger functions introduced in \S2 can be usefully expressed as
functional integrals

$$ \eqalign{ &S_{\s_1\cdots\s_n}(\x_1,\ldots,\x_n)
=\lim_{\b\to\io}\lim_{\L\to\infty}\cdot\cr\cdot& {\ig P(d\psi)
\,\,e^{-\l
\ig_{-L/2}^{L/2}\ig_0^\b
\psi^+_\x\psi^-_\x\psi^+_\h\psi^-_\h \delta(x_0-y_0)
v(x-y)d\x d\h}\,
\,\psi^{\s_1}_{\x_1}\cdots\psi^{\s_n}_{\x_n}\over \ig P(d\psi)
e^{-\l\ig_{-L/2}^{L/2}\ig_0^\b \cdots\cdots}}\cr}\Eq(3.1)$$
where the $\psi^\pm_\x$ are ``{\it Grassmanian variables}'', see for
instance sec. 3.1 in [BG95]. The
``integration'' with respect to $P(d\ps)$ is defined on monomials
$\ps^+_{\x_1}\ldots \ps^-_{\x_{2n}}$ by assigning the value of
$\media{\ps^+_{\x_1}\ldots \ps^-_{\x_{2n}}}\defi\ig
P(d\ps)\,\ps^+_{\x_1}\ldots \ps^-_{\x_{2n}}$ via Wick's rule with
propagators
$\media{\ps^-_\x\ps^-_\h}$ $=\media{\ps^+_{\x}\ps^+_{\h}}=0$
and
$$\eqalign{
&\media{\ps^-_{\x}\ps^+_{\x'}}=g(\x-\x')\cr
&g(\x)=\sum_{n\in Z^1,\,n_0\in Z^1}(-1)^{n_0}
\lis g(x_0+n_0, x+n L,\b)\ ,\cr
\lis g(\x)=&{1\over(2\pi)^{d+1}}\ig {e^{-ip_0 x_0-i p x}\over -ip_0+
({p}^2-p_F^2)/2} dp_0 d{p}\cr}\Eq(3.2)$$
so that $g(\x)$ is periodic in $x$ with period $L$ and antiperiodic in
$x_0$ with period $\b$.

The integration is extended linearly to even polynomials and to functions
of the Grassmanian fields that admit an entire even power series expansion,
like the exponential in \equ(3.1).

In this way the integrals in the numerator and in the denominator of
\equ(3.1) are defined as formal power series in $\l$. The series can
be quite easily shown to be convergent for $|\l|< O(L^{-1})$ and one of
the goals is to find conditions under which they can be
analytically continued to values of $\l$ which are, possibly, small but
independent of the size $L$ of the system.
\*

If we insist in fixing {\it a priori} the chemical potential the first
effect of the interaction will be that the singularity of the Fourier
transform of $\media{\pst_x^+\pst^-_0}$, see \equ(1.3), will no
longer be where it is located when $\l=0$, namely at $k=\pm p_F=\pm
\sqrt{2\m}$, but (reasonably) it will be shifted by $O(\l)$.

It is, therefore, more convenient to {\it keep the location of the
singularity fixed} at a prefixed value $p_F$. This can be achieved by
taking a $\l$--dependent chemical potential $\m=\fra12p_F^2+\n$, see \equ(2.1),
where $\n$ has to be conveniently chosen as a function of $\l$.

This means that instead of \equ(3.1) one studies the same expression
with the argument of the exponential modified into

$$\eqalign{
&-\n\ig_{-L/2}^{L/2}\ig_0^\b \psi^+_\x\psi^-_\x\,d\x
-\l\ig_{-L/2}^{L/2}\ig_0^\b
\psi^+_\x\psi^-_\x\psi^+_\h\psi^-_\h \delta(x_0-y_0)
v(x-y)d\x d\h\cr}\Eq(3.3)$$

{\it In a series of basic papers, [Lu60], [KL60], [LW60], Luttinger,
Kohn and Ward point out that this determines $\n$ as a power series in
$\l$ and it has the important effect of generating a power series for
the Schwinger functions which is finite to all orders, uniformly in
the size of the system}: had we fixed the chemical potential $\m$
rather than the Fermi momentum $p_F$ we would have obtained a power
series in $\l$ with coefficients diverging to all non trivial orders
as $L\to\io$.

Of course the latter divergence does not mean that the theory with a
fixed chemical potential cannot be defined: it simply means that such
a theory will have Schwinger functions with a singularity at a Fermi
momentum which is different from $\sqrt{2\m}$ so that the expansion
with reference to a free field with singularity at $\sqrt{2\m}$
contains diverging expressions. It makes also clear that it is likely
to be more convenient to develop a perturbation theory of the ground
state at fixed Fermi momentum rather than at fixed chemical potential.
Should one wish to study the problem at fixed chemical potential $\m$,
after developing the theory at fixed $p_F$ and obtaining the chemical
potential correction $\n$ in terms of $\l,p_F$, one can imagine, that
$\m=p_F+\n(p_F,\l)$ and solve this relation for $p_F$, as a function
of $\m,\l$
\*

A related important result due to Luttinger  ({\it ``Luttinger's theorem''})
states that {\it fixing the Fermi momentum is equivalent to fixing the
density which is $\r=\p p_F$} whether $\l=0$ or not: this was shown by
Luttinger, [Lu60], to hold to all orders of perturbation theory. A
mathematically rigorous proof would be desirable.
\*

\0{\bf\S4. The ultraviolet problem.}
\numsec=4\numfor=1\*

If $p_0^{-1}$  is the range of the interaction potential
and $\e(p)=(p^2-p_F^2)/2$
the propagator \equ(3.2) can be written as sum of two terms

$$\eqalign{
&\lis g(\x)= g^{(>0)}(\x)+ g^{(\le0)}(\x)\cr
&g^{(>0)}(\x)={1\over(2\pi)^{2}}\ig {(1-e^{-(q_0^2+
\e({q})^2)/p_0^4})\, e^{-i\k\cdot \x}\over -iq_0+ \e(q)} d\k\cr
&g^{(\le0)}(\x)=
{1\over(2\pi)^{d+1}}\ig {e^{-(q_0^2+
\e({q})^2)/p_0^4} \,e^{-i\k\cdot\x}\over -iq_0+ \e(q)} d\k\cr}
\Eq(4.1)$$
where $\x=(x_0,x)$, $\k=(q_0,q)$. The term $g^{(>0)}$ is the
``ultraviolet component of the propagator'' and $g^{(\le0)}$ is the
``infrared component''.

The decomposition can be used to introduce two auxiliary
``independent'' Grassmanian fields $\ps^{(>0)}_\x$ and $\ps^{(\le
0)}_\x$ so that $\ps_\x=\ps^{(>0)}_\x+\ps^{(\le 0)}_\x$. This means
that in evaluating the integrals in \equ(3.1) we can replace $\ps_x$
with $\ps^{(>0)}_\x+\ps^{(\le 0)}_\x$ and perform the Grassmanian
integration following the Wick's rule with propagators

$$ \media{\ps^{(\a)-}_\x\ps^{(\a)+}_\h}=g^{(\a)}(\x-\h)\qquad \a=
(>0),(\le0)\Eq(4.2)$$
while all the other propagators vanish. Calling $V(\ps)$ the
expression \equ(3.3) this is also written as the identity (``Fubini's
theorem'' for Grassmanian integrals)

$$\eqalign{
&\fra{\ig P(d\ps)\, e^{V(\ps)}
\ps^{\s_1}_{\x_1}\cdots}
{\ig P(d\ps)\, e^{V(\ps)}}\=\cr
&\=\fra{\ig P(d\ps^{(\le0)})P(d\ps^{(>0)})
\, e^{V(\ps^{(>0)}+\ps^{(\le0)})}
(\ps^{(>0)\s_1}_{\x_1}+\ps^{(\le0)\s_1}_{\x_1})\cdots}
{\ig P(d\ps)\, e^{V(\ps^{(>0)}+\ps^{(\le0)})}}\cr}\Eq(4.3)$$
\0{\it Remark:} attention should be paid to the fact that
\equ(4.1) deals with the ``infinite volume'' ($\b\to\io,L\to\io$)
limit $\lis g(\x)$ rather than with the $g(\x)$ of \equ(3.2). Hence
one should really deal with $g$ and decompose the latter into the sum
$g^{(>0)}(\x)+ g^{(\le0)}(\x)$. This generates a great variety of
``small'' problems both in the ultraviolet and, later, in the infrared
analysis. There is no few words way out of this (well known)
difficulty. Here we choose to ignore it except for a few necessary
comments when needed, because it is discussed widely in the
literature, see [BG95], [BM00a].

\*
The idea is to perform first the integral over the ``high frequency
part'' $\ps^{(>0)}$ of the field both in the numerator and in the
denominator of \equ(4.3). Fixing attention on the denominator (simpler
than the numerator which, however, can be treated in the same way) the
result will be written

$$\ig P(d\ps)\, e^{V(\ps)}=\ig P(d\ps^{(\le0)})
\,e^{\tilde V^{(0)}(\ps^{(\le0)})}
\Eq(4.4)$$
which in fact is a definition of $\tilde V^{(0)}$, the ``effective
potential on scale $p_0$''.

Technically there is a lot of work behind the latter relation: one has
to show that the result of the integration of the field $\ps^{(>0)}$
can be written in the form of an exponential of an effective potential
$\tilde V^{(0)}$: this means showing that the result can be expressed
as an exponential of

$$\eqalign{
&\tilde V^{(0)}(\ps^{(\le0)})=
\sum_{m\ge0,p\le m}^\io \ig \fra{d\x_1\ldots
d\x_{m}}{m!}  \tilde V^{(0)}_{m,p}(\x_1\ldots\x_{m})\cdot\cr
&\cdot\ps^{(\le0)+}_{\x_1}\ldots\ps^{(\le0)-}_{\x_p}
\dpr_{\x_{p+1}}\ps^{(\le0)+}_{\x_p}
\ldots \dpr_{\x_m}\ps^{(\le0)-}_{\x_{m}}\cr}\Eq(4.5)$$
and the kernels $\tilde V^{(0)}_{m,p}(\x_1\ldots\x_{m})$ are
\*

(i) {\it analytic in $\l,\n$}.
\*

(ii) {\it decay exponentially}
on scale $p_0$: which means that they are bounded by $C_n \exp- c p_0^{-1}
d(\x_1,\ldots\x_{2n})$ where $d(\x_1,\ldots\x_{2n})$ is the length
pf the shortest path connecting the points $\x_1,\ldots\x_{2n}$ in
$R^{2n}$ and $c>0, C_n$ are suitable constants.
\*

Note that already the $V(\ps)$, \cfr \equ(3.3) can
be cast in the form \equ(4.5).  Furthermore the Grassmanian
integral $P(d\ps)$ can be written as the ``Lebesgue Grassmanian
integral'' $d\ps^+d\ps^-$ times an exponential

$$P(d\ps)= const\, e^{-\ig_{-L/2}^{L/2}\ig_0^\b dx dx_0 \ps^+_\x
(\dpr_{x_0} -\fra12(\dpr_x-p_F^2))\ps^-_\x}\,d\ps^+d\ps^-\Eq(4.6)$$
where $d\ps^+d\ps^-$ is the Grassmanian integration with ``trivial
propagators', \ie with the only non vanishing propagator given by
$\media{\ps^-_\x\ps^+_\h}=\d(\x-\h)$. Therefore the argument of the
exponent can also be written in the form \equ(4.5) (involving only
derivatives of first order at most: as the second derivative can be
integrated by parts): a property that turns out to be quite important.
\*

\0{\it Remark:} the difficulty mentioned in the remark following
\equ(4.3) shows up very clearly here. The integrals in the exponential
in \equ(4.5) should be over the rectangle $[-L/2,L/2]\times[0,\b]$
with periodic--antiperiodic boundary conditions. However they are
extended to the whole $R^2$: this means that we have implicitly taken
the limits in \equ(3.1). However strictly speaking this does not make
sense unless we explain what it means to integrate over the whole
space a monomial in Grassmanian variables. The correct interpretation
is the following:

(a) keeping $\b,L$ finite and using the propagator $g(\x)$ in
\equ(3.2) one performs the integration over the ultraviolet components
$\ps^{(>0)}$ and one obtains \equ(4.4) with a $\tilde V^{(0)}$
similar to \equ(4.5) but with the coefficient kernels which now depend
on $L,\b$.

(b) the kernels converge to limits as $\b,L\to\io$ and the limits
$\tilde V^{(0)}_{m,p}(\x_1\ldots\x_{m})$ verify the properties stated after
\equ(4.5) (uniformly in $\b,L$).

(c) the identity \equ(4.4) means that if one wishes to compute
$$\fra{\ig P(d\ps^{(>0)})P(d\ps^{(\le 0)}) e^{V(\ps)}
\ps^{(\le0)\s_1}_{\x_1}\ps^{(\le0)\s_2}_{\x_2}\ldots}{
\ig P(d\ps^{(>0)})P(d\ps^{(\le 0)}) e^{V(\ps)}}\Eq(4.7)
$$
then one can ``simply'' compute

$$\fra{\ig P(d\ps^{(\le
0)}) e^{\tilde V^{(0)}(\ps^{(\le0)})}
\ps^{(\le0)\s_1}_{\x_1}\ps^{(\le0)\s_2}_{\x_2}\ldots}{
\ig P(d\ps^{(\le
0)}) e^{\tilde V^{(0)}(\ps^{(\le0)})}}\Eq(4.8)$$
by developing the
$e^{\tilde V^{(0)}}$ in powers of the fields and then apply Wick's
rule with propagator $g^{(\le0)}$ obtaining in this way a combination
of integrals of products of the kernels in \equ(4.5). And the series
converges.

(d) all divergences due to the infinite extension of the domains of
integration over the $\x$ variables disappear when one considers the
ratio in \equ(4.3) because of the fast decay of the kernels. As one
may suspect there will also be exchange of limits queries, which are
solved again by using the uniformity of the estimates.  See [BGPS94],
for instance, and [BM00a].

\*
Performing the ``remaining integral'' is therefore similar to the
original problem except that now the expression $\tilde V^{(0)}$ is
more involved but the propagator is simpler (being ``just''
$g^{(\le0)}$, \ie a propagator with an ultraviolet cut off at scale
$p_0^{-1}$).

In fact the decay of the coefficients $\tilde V^{(0)}_m$ make the
$\tilde V^{(0)}$ ``essentially local'' and the ``remaining integral''
is not really harder than the one obtained by replacing $\tilde
V^{(0)}$ by

$$\eqalign{&V^{(0)}(\ps^{(\le0)})=
\n_0\ig_{-L/2}^{L/2}\ig_0^\b \psi^{(\le0)+}_\x\psi^{(\le0)-}_\x\,d\x+\cr
&+\l_0 \ig_{-L/2}^{L/2}\ig_0^\b
\psi^{(\le0)+}_\x\psi^{(\le0)-}_\x\psi^{(\le0)+}_\h\psi^{(\le0)-}_\h\,
w(\x-\h)
\,d\x\,d\h \cr}\Eq(4.9)$$
where $w$ is a smooth potential with range  $p_0^{-1}$.

In fact in most treatments the above analysis is considered
``trivial'' and one just poses the problem of studying the ratio
of integrals

$$\fra{\ig P(d\ps^{(\le0)})
\, e^{-V^{(0)}(\ps^{(\le0)})}\,
\ps^{(\le0)\s_1}_{\x_1}\cdots}
{\ig P(d\ps^{(\le0)})\, e^{-V^{(0)}(\ps^{(\le0)})}}\Eq(4.10)$$
with $V^{(0)}$ given by \equ(4.9).
\*

{\it It should be noted that the ultraviolet problem does not even arise
when the fermions are supposed to be located on a lattice} because in
that case there is a minimum length scale, hence a maximum momentum.
\*

For this reason I concentrate here on the infrared problem, modeled by
\equ(4.10). However things are not so simple in the case of the
Luttinger model: in that model the inverse propagator diverges at
large momenta linearly rather than quadratically. This generates a non
trivial ultraviolet problem: it can be treated in a way analogous to
the above, but one has to exhibit various cancellations because the
expressions for the kernels of the effective potential $V^{(0)}$ are
given by apparently non convergent integrals. The analysis for the
(harder) Luttinger model case is carried out in detail in [GS93].
\*

\0{\bf \S5. The infrared problem and quasi particles.}
\numsec=5\numfor=1\*

The infrared problem (\ie understanding the large distance properties
of the Schwinger functions, or equivalently the singularities at
finite momentum of their Fourier transforms) is more interesting and
rich in structure. A naive application of perturbation theory leads to
facing the fact that the propagators oscillate on scale $p_F^{-1}$ and
decay slowly at infinity.

Quasi particles arise when one attempts to disentangle the
oscillations and the decay at $\io$. Technically one remarks that the
propagator for the functional integrals with
respect to $P(d\ps)$ can be written, setting $\x=(t,\xx)$ and
$\k=(k_0,\kk)$ in general dimension $d\ge1$, as

$$\eqalign{
g(\x)=&\ig{d^{d+1} q\over(2\p)^{d+1}}{e^{-i(q_0t+\V q\cdot\xx)}
\over-iq_0+({\V q}^2-p^2_F)/2m}=\ig d\oo e^{-ip_F\oo\xx}g(\xx,t,\oo)=\cr
=&\ig e^{-ip_F\oo\xx} \,d\oo\ig \fra{
e^{-i k_0 t+\kk\cdot \xx}}{\e(\kk,\oo)}\fra{d^{d+1}\k}{(2\p)^{d+1}}
\cr}\Eq(5.1)$$
where $\oo$ is a unit vector and $d\oo$ is the integration over the
unit sphere in $R^d$ normalized to $1$ (if the dimension $d=1$
integrating over $\oo$ means $2^{-1}\sum_{\o=\pm}\cdot$). An
elementary calculation shows that (at least if $d$ is odd)
$g(\xx,t,\oo)$, {\it which is not unique}, can be defined so that
$\e(k,\oo)=-ik_0+v_F\oo\cdot\kk+O(\kk^2)$ (see
\S5 and appendix A in [BG90]).
\*

This means that the free fermion system in a ground state with Fermi
momentum at $p_F$ can be considered as a system of {\it ``quasi
particles''} in the vacuum carrying an ``{\it intrinsic}'' linear
momentum equal to a Fermi sphere momentum $p_F\oo$ in addition to the
``{\it external}'' momentum $\kk$. The {\it dispersion relation} is
{\it almost} linear in the sense that the system on large scales, \ie
$\k$ small, will show a dispersion relation essentially identical to
$\e(\kk,\oo)=\oo\cdot\kk\,v_F$: this property seems to remain valid
even in presence of interaction so that the intuition can be led by
the idea that the {\it quasi particles} must be taken seriously, see
sec. 5 in [BG90]. Formally \equ(5.1) can be regarded, \cfr also
\equ(1.2), as the propagator of a composite field defined as
$$\ps^\pm_{\V x,t}=\ig d\oo e^{\pm ip_F\oo\V x}
\ps^\pm_{\V x,t,\V\o}\Eq(5.2)
$$
with the fields $\ps^\pm_{\V x,t,\V\o}$ being Grassmanian fields with
propagators $g(\x,\oo)$.
\*

\0{\it Remark:} 
It is important to stress that the above fields are just Grassmanian
fields rather than the usual fermionic fields. If we wanted, instead,
to think of the quasi particles fields as Grassmanian fields which
correspond to {\it physical particles} {\it it would be necessary that
the propagators $g(\x,\oo)$ have the ``reflection positivity''
property}, [Si74]: and this would also be a criterion to fix the
arbitrariness in the choice of the representation \equ(5.1) mentioned
above. However it is not known whether such a choice is possible. The
success of the quasi particles--based view of the theory of the ground
states of fermions leads us to feel that this might be possible.
\*

It is difficult to give a meaning to the simple and captivating
{\it approximation}

$$g(\x,\oo)=\ig \fra{d^{d+1}\k}{(2\p)^{d+1}}
\fra{e^{-i(k_0t+\kk\xx)}
}{-ik_0+v_F \oo\cdot \kk}\Eq(5.3)$$
if $d>1$ because it corresponds to a system of fermions with a linear
dispersion relation which {\it therefore} will give rise to
ultraviolet instabilities: possibly when the potential is repulsive
and certainly when it is attractive. Note, however, that it makes
sense if $d=1$ because in this case the system is stable {\it even}
with a linear kinetic energy: in fact it becomes almost precisely the
Luttinger model (it differs from it because of the presence of
``extra'' cross terms in the interaction, \cfr comment following
\equ(1.8)).

However \equ(5.3) gives a good representation of the propagators {\it for
small $\k$}. Therefore the approximation can be expected to be
reasonable in the sense that its version with an ultraviolet cut--off, \ie

$$g^{(\le 0)}(\x,\oo)=\ig \fra{d^{d+1}\k}{(2\p)^{d+1}}
e^{-i(k_0t+\kk\xx)}
\fra{e^{-p_0^{-4}(k_0^2+v_F^2\kk^2)}}{-ik_0+v_F \oo\cdot \kk},\Eq(5.4)$$
can be used to study the infrared problems presented by the \equ(4.9),
\equ(4.10).
\*

The papers [BG90] and [BGPS94] consider the integrals in \equ(4.10)
with a propagator \equ(5.4) and an interaction $V^{(0)}$ even simpler
than the one in \equ(4.9), namely
$$\eqalign{
&V^{(0)}(\ps^{(\le0)})=
\n_0\ig_{-L/2}^{L/2}\ig_0^\b \sum_{\o=\pm}
\psi^{(\le0)+}_{\o,\x}\psi^{(\le0)-}_{\o,\x}\,d\x+\cr
&+\l_0 \ig_{-L/2}^{L/2}\ig_0^\b
\psi^{(\le0)+}_{+,\x}\psi^{(\le0)-}_{+,\x}
\psi^{(\le0)+}_{-,\x}\psi^{(\le0)-}_{-,\x}
\,d\x \cr}\Eq(5.5)$$
where several crossed terms have been eliminated and the interaction
has been made strictly local (which, in the spinless case, is possible
only if we think of the system as made with quasi particles because
the exclusion principle makes local interactions which are quartic in
spinless fermion fields vanish identically).
\*

In the quoted papers it is shown in detail that if $d=1$ the
understanding of the integrals in \equ(4.10) with a propagator
\equ(5.4) and an interaction $V^{(0)}$ given by \equ(5.5) suffice to
solve in a mathematically complete way the problem of the ground state
of \equ(1.1) and to obtain Tomonaga's main result that the anomaly at
the Fermi surface is not $0$ and in fact it is an analytic function of
the coupling $\l$. Of course this is no surprise because of the works
[Lu63], [ML65].
\*

{\it Why to redo Tomonaga's work in mathematically precise way?} the
point is that the notion of ``understanding'' and of ``proof'' of a
physical result evolve. And many of the results that were considered
established at the time ($\sim1950$) have come under scrutiny and have
been cast into a more rigorous form. This is necessary mostly because
the attempts to extend them to other cases (namely higher dimension or
even one dimension with spin, or just with more structure, see
question (2) in \S8) have failed and therefore it becomes necessary to
have a clear idea of what really fails and what can still be usefully
taken over to attack harder problems. This necessity is similar to the
need that became evident in the 1960's to have a more rigorous
foundation of the theory of ensembles in equilibrium statistical
mechanics. It should be clear that revisiting Tomonaga's theory in no
way implies that there are faults in the original work: it simply did
not deal with questions that at the time were, rightly, not considered
important.
\*
\1

\0{\bf\S6. Renormalization group and the infrared problem.}
\numsec=6\numfor=1\*

We consider the problem of computing the ``partition function'', \ie
the problem of studying the integral in the denominator of \equ(4.9)
with a propagator given by \equ(5.4) and an interaction given by
\equ(5.5). We define

$$g^{(h)}(\x,\o)\defi \ig \fra{d\k}{(2\p)^2}
\fra{\big(e^{-2^{-2h} \k^2} -e^{-2^{-2(h-1) } \k^2}\big)}{-i
k_0+v_F\,\o\, k} e^{i(k_0t+\o\, k x)}=2^h g^{(0)}(2^h\x,\o)\Eq(6.1)$$
$h=0,-1,-2,\ldots$, $\o=\pm$, $\k=(k_0,k)$, $\k^2=(k_0^2+v_F^2
k^2)/p_0^4$. Therefore if $Z_0\defi 1$

$$\fra1{Z_0} g^{(\le0)}(\x,\o)=\fra1{Z_0} \sum_{h=0}^{-\io}
2^h\,g^{(h)}(2^h\x,\o)\defi \fra1{Z_0} g^{(0)}(\x,\o)+
\fra1{Z_0} g^{(\le-1)}(\x,\o)\Eq(6.2)$$
and the integration \equ(4.10) can be thought of as an integration
over quasi particle fields $\ps^{(\le0)}_{\o,\x}$ which are
decomposable as the sum of two fields $\ps^{(0)}_{\o,\x}$ and
$\ps^{(\le-1)}_{\o,\x}$ with propagators $\fra1{Z_0} g^{(0)}(\x,\o)$
and $\fra1{Z_0} g^{(\le-1)}(\x,\o)$:

$$\ig P(d\ps^{(\le0)}) e^{V^{(0)}(\sqrt{Z_0}\ps)}\=
\ig P_{Z_0}(d\ps^{(0)}) P_{Z_0}(d\ps^{(\le-1)})
e^{V^{(0)}(\sqrt{Z_0}(\ps^{(0)}+\ps^{(\le -1)}))}\Eq(6.3)$$
This is convenient because ``we know'' how to compute the integral
over $\ps^{(0)}$ by perturbation theory.  This means that the
techniques to study the integral although not trivial are,
nevertheless, well established mainly via the results in [Le87] which
provide us with a technique designed to take advamtage of the
fermionic nature of the fields. The result is the expected one: one
can basically compute to second order of perturbation in $\l$ and
neglect the rest. More precisely one can prove that the result of the
integration (in the sense discussed in the remark following \equ(4.5))
is

$$\eqalign{
&\ig  P_{Z_0}(d\ps^{(\le-1)})e^{\lis
V^{(0)}(\sqrt{Z_0}\ps^{(\le-1)})}, \qquad{\rm with}\qquad
\lis V^{(0)}(\sqrt{Z_0}\ps^{(\le-1)})=\cr
&=\sum_{n\ge 0, \a} \ig
\sqrt{Z_0}^n V_\a(\x_1,\ldots,\x_n) \F_{\x_1}\cdots
\F_{\x_n}\, e^{ip_F\sum_i\s_i\oo_i x_i}\, \fra{d\x}{n!}\cr}\Eq(6.4)$$
where $\F_{\x_i}$ is either $\ps^{\s_i}_{\x_i}$, $\s_i=\pm$, or a
derivative of this field; and $\a$ denotes the labels $\s_i$ as well
as the labels necessary to identify which of the fields $\F$ are
differentiated and which are not; we also suppose that the $\ps^+$
fields are to the left of the $\ps^-$ (recall, however, that the
fields are Grassmanian so that they anticommute). The number $n$ must
be, obviously, even.
\*

The kernels $V_\a(\x_1,\ldots,\x_n)$ are analytic functions of
$\l_0,\n_0$ convergent with a convergence radius which {\it is
independent of the sizes $L,\b$ of the system} and decay
exponentially fast on scale $p_0^{-1}$, in the sense following
\equ(4.5): the technique for this proof is in [Le87] and the analysis
can be found in [BGPS94].
\*

It would not be wise to simply iterate the procedure calling
$V^{(-1)}$ the ``effective interaction'' $\lis V^{(0)}$ and
(with the procedure already used in the previous splitting, \cfr
\equ(6.2))  splitting the field $\ps^{(\le -1)}$ into $\ps^{(-1)}+\ps^{(\le
-2)}$: because the result, under further iteration, would be {\it a
progressive worsening} of the bounds on the constants and a decrease
in the convergence radius of the expansions in powers of
$]l_0,\n_0$. This is basically due to the fact that $\n_0$ {\it cannot
be arbitrary} under the only condition that it is small: {\it it has to be
tuned} so that the Schwinger functions have singularities at momentum
$\pm p_F$.

The correct procedure will be to adjust recursively the value of
$\n_0$ as a function of $\l_0$ {\it keeping the singularity at fixed
momentum}.  For this purpose one distinguishes the {\it relevant} part
of the interaction $\lis V^{(0)}$ from the {\it irrelevant} part. The latter
is {\it not} small (in a sense it is the most important part!): the
relevant part is just a part of $\lis V^{(0)}$ whose absence would allow
us to perform a naive recursion without loss in the size of the
constants or of the convergence radii.

The identification of the relevant terms is a matter of dimensional
analysis, at least in the simplest cases, see [Ga85], [BG95]. What we
call here ``relevant'' is a set of terms (see below) which in the
usual nomenclature of the renormalization group approaches is further
divided into relevant and marginal terms. In the present case the
relevant part is the ``{\it local part}'' of the terms which are
quadratic and quartic in the fields. If $\d_{a,b}$ denotes a Kronecker
delta such local part is defined by

$$\eqalign{
&\LL \psi_{\x_1{\o}_1}^+ \psi_{\x_2{\o}_2}^+
\psi_{\x_3{\o}_3}^-\psi_{\x_4{\o}_4}^-\ =
{\d_{\o_1+\o_2+\o_3+\o_4,0}}\,\fra12 \sum_{j=1}^2
\psi_{\x_j{\o}_1}^+ \psi_{\x_j{\o}_2}^+
\psi_{\x_j{\o}_3}^-\psi_{\x_j{\o}_4}^-\ \cr
&\LL \psi_{\x_1{\o}_1}^+\psi_{\x_2{\o}_2}^-\ =
\d_{\o_1,\o_2}(\psi_{\x_1{\omega}_1}^+ \psi_{\x_2{\omega}_2}^- +
\psi_{\x_1 {\omega}_1}^+ (\x_2-\x_1)\cdot
D_\x\, \psi_{\x_2 {\omega}_2}^-)
\cr}\Eq(6.5)$$
where $D_\x = (\dpr_t, \,\dpr_x)$.

Applying the operator $\LL$ to the expression in \equ(6.4) one
realizes that the result can be espressed as a linear combination of
the following Grassmanian monomials
$$\eqalignno{
F_1 =  & -\ig \psi^+_{+,\x}\psi^-_{+,\x}\psi^+_{-,\x}\psi^-_{-,\x} d\x\cr
F_2 =  & -\ig \sum_{\o=\pm}\psi^+_{{\o},\x}\psi^-_{{\o},\x}
         \,d\x\ ,&\eq(6.6)\cr
F_3 =  & -\ig \sum_{\o=\pm}\psi^+_{{\o},\x}
         (-i\,v_F\,\o\dpr_x)\psi^-_{{\o},\x} \,d\x\ ,\cr
F_4 =  & -\ig  \sum_{\o=\pm}\psi^+_{{\o},x} \partial_t \psi^-_{x{\o},x}
          \,d\x\cr}$$
Note that there is only one possible non zero local term of fourth
order in the field because of the Fermi statistics and the facts that
$\o=\pm 1$ and that our fermions are spinless. Note also that the
above localization operation would be completely different ({\it and
useless}) if we had not introduced the quasi particles: for instance
$F_1$ would simply vanish because of the Fermi statistics as it would
involve the square of Fermi fields.
\*

Applying the operator $\LL$ to $\lis V^{(0)}(\sqrt{Z_{0}}\ps^{(\le
-1)})$ one obtains an expression 

$$\eqalign{
&\,\lis \l^{(0)}\,Z_0^2 F_1(\ps^{(\le-1)})+
\lis \n^{(0)} Z_0 F_2(\ps^{(\le-1)})+\cr
&+\lis\a^{(0)}Z_0 F_3(\ps^{(\le-1)})+ \lis \z^{(0)}
Z_0 F_4(\ps^{(\le-1)})+(1-\LL) \lis V^{(0)}\cr}\Eq(6.7)$$
where $\lis \l^{(0)},\lis \n^{(0)},\lis\a^{(0)},\lis \z^{(0)}$ are
simple combinations of integrals of the kernels, see \equ(6.4), for
the monomials quadratic or quartic in the fields and are therefore
analytic in $\l_0,\n_0$.
\*

The simplest way to proceed is to consider a representation of
$P_{Z_0}(d\ps)$ analogous to \equ(4.6) for the integrations over the
quasi particle fields, with the quadratic form $\ps^{(\le-1)+}_{\x}
(\dpr_t +(-\fra12\dpr_x^2-\fra12 p_F^2))\ps^{(\le-1)-}_{\x}$ replaced
by $Z_0\sum_{\o=\pm}\ps^{(\le-1)+}_{\o,\x} (\dpr_t
-iv_F\o\dpr_x)\G_{-1}(\Dpr)\ps^{(\le-1)-}_{\o,\x}$ where $\G_h(\Dpr)$ is
the operator which multiplies the Fourier transforms by $\G_h(\k)=
e^{+(2^hp_0)^{-2}(k_0^2+v_F^2k^2)}$: indeed this choice attributes to the
field $\ps^{(\le-1)-}_{\o,\x}$ the correct propagator
$g^{(\le-1)}(\x)$, \cfr \equ(6.2). Then the integral to be performed
looks like

$$\eqalign{ &const\, \ig e^{-Z_0 \sum_{\o=\pm}\ig
\ps^{(\le-1)+}_{\o,\x} (\dpr_t -iv_F\o\dpr)
\G_{-1}(\V\dpr)\ps^{(\le-1)-}_{\o,\x}d\x}\cdot\cr & \cdot e^{\lis\n_0Z_0
F_{2}(\ps^{(\le-1)})+\lis
\l^{(0)}\,Z_0^2 F_1(\ps^{(\le-1)}) +(1-\LL) \lis
V^{(0)}}\cdot\cr
&\cdot e^{\lis\a^{(0)}Z_0
F_3(\ps^{(\le-1)})+ \lis \z^{(0)} Z_0 F_4(\ps^{(\le-1)})}\,d\ps^+d\ps^-
\cr}\Eq(6.8)$$
and it is natural to collect the quadratic parts defining

$$\eqalign{
&Z_{-1}\defi Z_0+ \lis\a_0 Z_0,\qquad
\z_{-1} Z_{-1}\defi (\lis \z^{(0)}-\lis\a_0)Z_0 \cr
&\l_{-1}Z_{-1}^2=  \lis \l^{(0)}Z_0^2,\qquad
\n_{-1} Z_{-1}= 2\, \lis \n_0Z_0\cr}\Eq(6.9)$$
Here the factor $2$ in front of the ``relevant coupling'' $\lis\n_0$is
natural: if the theory is developed only to first order (a deed called
``power counting'' in the jargon of the renormalization group, see
eq. (6.18) in [BG95]) the result is that $Z_{-1}=Z_0$ and
$\n_{-1}=2\n_0$. This transforms the integral into

$$\eqalign{
\ig P_{Z_{-1}}(d\ps^{(\le-1)})&\, e^{ Z_{-1} \z_{-1}
F_4(\ps^{(\le-1)})+Z_{-1} \n_{-1}
F_2(\ps^{(\le-1)})}\cdot\cr
&\cdot e^{Z_{-1}^2 \l_{-1} F_1(\ps^{(\le-1)})+
W_h(\sqrt{Z_{-1}}\ps^{(\le-1)})}\cr}\Eq(6.10)$$
The procedure can now be iterated and the integrals can be performed
in the same way defining recursively $\l_h,\n_h,\z_h$ called ``{\it
running couplings}'', $Z_h$ called ``{\it wave function
renormalization}'' and $W_h$ called ``{\it irrelevant operators}'' for
$h=0,-1,-2\ldots$. The ratio $Z_h/Z_{h+1}$ will be called the {\it
``wave function renormalization rate''}.
\*

\0{\it Remarks:}
(i) Note that
$$\eqalign{
&W_h(\sqrt{Z_{-1}}(\ps^{(\le-1)}))=(1-\LL)
\lis V^{(0)}(\sqrt{Z_{0}}\ps^{(\le -1)})+\cr
&+\lis\a_0 Z_0\ig \ps^{(\le-1)+}_\x (\dpr_t-iv_F\o\dpr_x)
(1-\G_{-1}(\Dpr))\ps^{(\le-1)-}_\x\,d\x\cr}\Eq(6.11)$$
(ii) In [BG90], [BGPS94], [BG95] the definitions of $W_h$ and/or of the
localization operator $\LL$ are slightly different: more involved but,
possibly, more convenient for performing estimates. The choices
differ, however, by ``irrelevant terms'' and are (therefore)
equivalent.
\*

{\it The basic bound is that the running couplings on scale $h$ and
the kernels that are coefficients in the definition of $W_h$ are {\it
analytic} in $\V v_{h'}=(\l_{h'},\n_{h'},\z_{h'})$ for $h'>h$ under
the condition that all the latter quantities and $|Z_{h'}/Z_{h'1}-1|$
for $h'>h$ are small enough independently on the size $L$ of the
system and also on the scale $h$.}
\*

The relation between the running couplings can be written, therefore,
as

$$\eqalign{
\V v_{h-1}=&\ M_h (\V v_h+ \lis B_h(\V v_h,\V v_{h+1},\ldots,\V v_0;
\fra{Z_{h}}{Z_{h+1}},\ldots)\ ,\cr
1=&\ \fra{Z_h}{Z_{h-1}}\,\big(1+\lis A_h(\V v_h,\V v_{h+1},\ldots,\V v_0;
\fra{Z_{h}}{Z_{h+1}},\ldots)\big)\ ,\cr}\Eq(6.12)$$
where $M_h$ is a diagonal matrix with matrix elements $(Z_h/Z_{h-1})^2$,
$2\,Z_h/Z_{h-1}$, $Z_h/Z_{h-1}$ and the functions $\lis B_h,\lis A_h$ are
analytic under the conditions that the running couplings and the
renormalization rate are small enough independently of $h\le0$.

Furthermore one can prove, [BG90], [BGPS94], that after eliminating
the wave function renormalization rates in the first of \equ(6.12) by
using recursively the second equation the relation
\equ(6.12) can be written

$$\eqalign{ &\fra{Z_{h}}{Z_{h+1}}= 1+ B'_h(\V v_{h+1},\V v_h,\ldots,\V
v_0)\cr &\V v_h=\L\,\V v_{h+1} + \V B_h(\V v_{h+1},\V v_h,\ldots,
\V v_0)\cr}\Eq(6.13)$$
with $B'_h,\V B_h$ analytic in their arguments $\V v_{h'}$, under the
conditions that the running couplings are small enough independently
of $h\le0$, and $\L$ is the diagonal matrix with diagonal $(1,1,2)$.
\*

At this point the strategy is clear: the \equ(6.13), called the {\it
beta functional}, defines a ``flow'', called the {\it renormalization
flow}, in the space of the running couplings $\V v=(\l,\n,\z)$ which
is well defined as long as the running couplings themselves stay small
enough in the above sense so that perturbation theory can be applied
to pass from one scale to the next (\ie as long as $|\V v_h|$ is so
small to be inside the convergence domain of the series expressing the
functions $B'_h,\V B_h$). The initial values are $\l_0,\n_0,\z_0$:
$\l_0$ will have to be taken small (so that the perturbation theory
can be applied at least to perform the first integration, \ie the
integration over $\ps^{(0)}$), $\n_0$ also will have to be taken small
and $\z_0$ {\it must} be taken $\z_0=0$ (because we start the analysis
from \equ(5.5) which contains no term like $F_4$, see \equ(6.6.).

Note that if we studied the problem without the approximation
introduced after \equ(5.5) (starting from the Hamiltonian \equ(3.3))
we would end up after the ultraviolet integration with a $V^{(0)}$
which is much more involved than the expression \equ(5.5) with which
we began the infrared analysis. Therefore the localization operator
acting on $V^{(0)}$ may (as indeed it does) produce also a term of the
type $F_4$: which implies that $\z_0$ will have a non zero value. This
however does not help to provide us with more freedom because this
$\z_0$ is {\it not arbitrary} being a function of the couplings
$\l,\n$ in the original Hamiltonian \equ(3.3): {\it it is intrinsic to
the model that there are only two adjustable constants in the
Hamiltonian}.
\*

\0{\bf\S7. The vanishing of the beta function: the role of the
Luttinger model.}
\numsec=7\numfor=1\*

The only freedom that we have to ``make things work'' is the choice
$\l_0$ (or of $\l$ in the theory without approximations) small
``enough'' and the selection of $\n_0$ (or of $\n$ in the theory
without approximations).

What has been sketched until now is technically involved, but it is
not really difficult because the techniques for a rigorous multiscale
analysis has been established since a long time.  In many problems,
among which the $d$--dimensional Fermi systems, see [BG90], or the
scalar field theories, see [Ga85], [BG95], {\it it is, in a sense, a
pure matter of technical routine work to prove that the successive
integration of the components of the fields on various scales can be
reduced to the study of a renormalization flow like
\equ(6.12), under the assumption that the running constants stay small
enough.}

In itself such a work {\it does not} really mean much unless one is
able to exhibit a trajectory of the flow that keeps the running
constants small enough to allow us to apply perturbation theory to
compute the effective potentials from one scale to the next via
convergent series for the kernels coefficients.
\*

\0{\it Remarks:} (1) It is indeed very easy to work out alternative, 
substantially different, integration algorithms which work as
beautifully as the one described above {\it but which are completely
useless}: a nice and relevant example is obtained by keeping $Z_h\=1$
and, correspondingly, not eliminating the quadratic terms with
coefficients $\lis \a_h$ that are generated at each integration of a
field component $\ps^{(h)}$. Everything works fine: {\it however one
can prove that no initial data $\l_0,\n_0$ exist which keep the
constants $\l_h,\n_h,\a_h,\z_h$ small for all $h$ if $\a_0=0,\z_0=0$
(or, in the theory without approximations, if $\l$ and $\n$ are the
only free parameters )!}

(2) The only interest of the latter remark (1) lies in the fact that if it
had been possible to find $\n_0$ as a function of $\l_0$ so that
$\l_h,\n_h,\a_h,z_h$ stayed small for all $h\le0$ then it would follow
(with some extra work, see [BGPS94]) that the singularity of the one
particle Schwinger function at the Fermi surface would be a
discontinuity, {\it as in the non interacting case}. This failure
shows that the latter property cannot be derived as a consequence of a
perturbation analysis, if true (it is not! as we know since Tomonaga's
work). 

(3) Of course the impossibility of a normal Fermi surface in generic
$1$--dimensional many fermions systems has been known since the early
days of many body theory as a consequence of the divergence of the
second order corrections to the Schwinger functions (divergence of the
``self--energy'' which occurs in dimension $1$).
\*

Therefore {\it after} the above (long and tiring) set up of the
renormalization flow the real work starts.

It was well known that there were interesting conjectures about a
number of cancellations that occur in the theory of one dimensional
spinless fermions. In our language they can be summarized by saying
that ``the beta function effectively vanishes'', see [So79].

It is very difficult however to find a proof by computing the beta
functional to all orders.  Introducing the
function, called the ``beta function'',

$$\V \b(\V v)\defi \lim_{h\to-\io} \V B_h(\V v,\V v,\ldots,\V
v), \qquad \V v=(\l,\n,\z)\in R^3\Eq(7.1)$$
a major simplification occurs by realizing, see [BG90], [BGPS94], that
{\it there is a solution to the flow generated by the recursion
\equ(6.13) with $|\V v_h|$ so small that all running couplings stay
within the convergence domain of the functions $\V B_h$ provided the
components (relative to the contants $\l$ and $\z$) $\b_1,\b_2$ of the
beta function $\V\b=(\b_1,\b_2,\b_3)$ vanish.}

\*
A proof of the latter property looks like an easier problem. However
the first proof came from an indirect argument that we describe below.

(1) one first repeats the above analysis for the Luttinger model: this
    was assumed possible in [BG90], [BGM92], and later proved in
    [GS93]: the difficulty being essentially in the treatment of the
    ultraviolet problem.

(2) once the beta functional has been defined for the Luttinger model
    it has been remarked in [BG90] that although the beta functionals
    of the Luttinger model and the corresponding one of the
    ``realistic'' model in \equ(1.1) are different the beta functions of
    the two models {\it coincide}, [BG90].

(3) one checks that if $\b_1$ or $\b_2$ did not vanish this would
    contradict the exact solution of the Luttinger model, [BG90],
    [BGM92], [BGPF94].

This allowed to prove the existence of a trajectory of the
renormalization flow which stays close to the origin and within the
radius of convergence of all the expansions in the running couplings
considered (including that of the beta functional) thus completing
the theory of the $1$--dimensional spinless Fermni gas at small
coupling yielding also the analyticity of the anomaly exponent in
$\l$.

The third component of the beta function does not vanish: this is
however not necessary. In fact the constant $\n_h$ tends to diverge
and very fast ($O(2^{-h})$): but this can be easily counterbalanced
because the initial value of $\n_0$ is free and we can tune it so that
$\n_h$ does not diverge. In fact there can be only one value of
$\n_0$ which has this property (essentially because, near a hyperbolic
equilibrium attractive or marginal along the $\l$--axis and repulsive
along the $\n$--axis, given $\l$ small enough one can find only one
value of $\n$ small such that the evolution of the datum $\l,\n$ stays
close to the origin forever).

\*

\0{\bf\S8. Outlook.}
\numsec=8\numfor=1\*

One can wonder whether a simpler symmetry argument can be given to
prove that the beta function $\b_1,\b_2$ components vanish: after
[BG90] this has been attempted in [MD93] who however neglect the
(necessary) presence of the decreasing cut--off $2^h p_0$, for
$h\to-\io$, arising in the successive integrations of the infrared
scales.

Nevertheless such a proof must be possible and recently new attempts
at finding it are being considered, [BM00b]: this would be a major
achievement which, however, would not diminish the importance of the
idea of Luttinger to enucleate out of Tomonaga's model a much simpler
model that really catches all its main features. The existence of the
indirect proof of the vanishing of the beta function based in an
essential way on the [ML65] solution of the Luttinger model, is an
important factor in the search of the proof based on symmetry
considerations.
\*

Mainly through the work of Mastropietro and collaborators the
Luttinger model (together with its extension by Mattis, [Ma64]) has
received several other applications in the frame of the
renormalization group approach to the ground state of Fermi systems;
for a complete review see [GM00] where, among other results, the
interesting phenomenon of the $\l$ dependence of the anomalies in the
higher Schwinger functions is summarized as well as the results of the
theory of one dimensional fermions in a periodic or {\it quasi
periodic} potential (with Fermi momentum $p_F$ in the bands or at the
top of a band).
\*

Another somewhat unexpected development has been the determination of
the asymptotic behavior of the correlation functions in eight vertex
models or $XYZ$ models in a magnetic field for values of the
parameters that do not correspond to exactly soluble points, [Ma99b],
[BM00a]. A corollary of the latter papers should be the determination
of the critical exponents of Ising type models even with non nearest
neighbour interactions, see [Ma99b]. Applying the remark in [Sp99]
concerning the translation of the next neighbor $2$--dimensional Ising
model into a fermionic spinless $1$--dimensional problem it could lead
to a strong extension of the theory of the critical point in such
model covering a wide range of other models which (unlike the case
considered in [Sp99]) show non universal critical exponents, see also
remarks to sec. 1.6 of [BM00a], [Ma00] and the review
[GM00]. Furthermore the technique used in [Ma99b], [BM00a] is quite
different from that in [Sp99].
\*

I conclude by mentioning a few open problems for small coupling:
\*

(1) spinning fermions on a line with attractive interaction. See
    [BM95] for the repulsive case

(2) spinless fermions on two parallel lines with attractive
    interaction, see [Ma99a] for the repulsive case.

(3) two and three dimensional Fermi surface properties(!).

(4) a mathematical proof of the identity $\r=\p p_F$ between the
    density $\r$ and the Fermi momentum $p_F$, see [Lu60] and \S3
    above, at small coupling $\l$.

\0and it is, perhaps, surprising to find the first two in a list of
open problems, while (3) is well known to require new ideas to be
understoon and (4) seems understandable with the techniques that we
know.
\*

We can say that the work of Luttinger [Lu63] proposing the exactly
soluble model bearing his name and providing some of the ideas that
contributed to its exact solution in [ML65] has been, and remains, a
landmark in Condensed Matter Physics (with its clarifying function of
Tomonaga's work) and in Mathematical Physics (as an example of how to
build a model that catches all the relevant features of a realistic
model and, yet, it is more tractable) and it has shed new light on the
difficult subject of the theory of low temperature quantum systems.
\*\*

\0{\bf References}
\*
{\parskip=1truemm\nota

\0{\bf[BG90] \sl Benfatto, G., Gallavotti, G.:} {\it Perturbation
theory of the Fermi surface in a quantum liquid. A general quasi
particle formalism and one dimensional systems}, J. Statistical
Physics, {\bf59}, 541-660, 1990. See also [BGPS94].

\0{\bf[BG95] \sl Benfatto, G., Gallavotti, G.:} {\it Renormalization
group}, p. 1--143, Princeton University Press, 1995.

\0{\bf [BGM92] \sl Benfatto, G., Gallavotti, G., Mastropietro, V.:}
{\it Renormalization group and the Fermi surface in the Luttinger
model}, Physical Review {\bf B 45}, 5468- 5480, 1992.

\0{\bf[BGPS94] \sl Benfatto, G., Gallavotti, G., Procacci, A.,
Scoppola, B.:} {\it Beta function and Schwinger functions for many
fermion systems in one dimension. Anomaly of the Fermi surface},
Communications in Mathematical Physics {\bf 160}, 93--172, 1994.

\0{\bf[BM00a] \sl Benfatto, G., Mastropietro, V.:} {\it
Renormalization group, hidden symmetries and approximate Ward
identities in the XYZ model,} I, II, in print on Communications in
Mathematical Physics, preprint 2000, in {\tt
http://ipparco. roma1.infn.it}.

\0{\bf[BM00b] \sl Benfatto, G., Mastropietro, V.:} private communication.

\0{\bf[BM95] \sl Bonetto, F., Mastropietro, V.}  {\it Beta function
and anomaly of the Fermi surface for a $d=1$ system of interacting
fermions in a periodic potential}, Communications in Mathematical
Physics, {\bf172}, 57--93, 1995.

\0{\bf[FT90] \sl Feldman, J., Trubowitz, E.:} {\it Perturbation theory
for many fermion systems}, Helvetica Physica Acta {\bf 63}, 157--260,
1990. And: {\it The flow of an electron-positron system to the
superconducting state}, Helvetica Physica Acta {\bf 64}, 213--357,
1991.

\0{\bf[Ga85] \sl Gallavotti, G.:} {\it Renormalization Theory and
Ultraviolet Stability for Scalar Fields via Renormalization Group
Methods}, Reviews of Modern Physics, {\bf 57}, 471--562, 1985.

\0{\bf[GM00] \sl Gentile, G., Mastropietro, V.:} {\it Renormalization
Group for One-Dimen\-sional Fermions: a Review on Mathematical
Results}, preprint, {\tt http://ipparco. roma1.infn.it}, in print in
Physics reports.

\0{\bf[GS93]: \sl Gentile, G., Scoppola, B.:} {\it Renormalization
group and the ultraviolet problem in the Luttinger model},
Communications in Mathematical Physics, {\bf 154}, 135--179, 1993. And
{\sl Gentile, G.}, Thesis, 1991, unpublished.

\0{\bf[KL60] \sl Kohn, W., Luttinger, J.M.:} {\it Ground state energy
of a many fermion system}, Physical Review {\bf 118}, 41--45, 1960.

\0{\bf[KL73] \sl Kac, M., Luttinger, J.M.:} {\it Bose--Einstein
condensation in the presence of impurities} $I$ and $II$, Journal of
athematical Physics, {\bf14}, 1626--1630, 1973, and {\bf15}, 183--186,
1974.

\0{\bf[KLY88] \sl Kennedy, T., Lieb, E., Shastry, ?.}: {\it The $XY$
model has long range order for all spins and all dimensions greater
than one}, Physical Review Letters, {\bf 61}, 2582--2584, 1988.

\0{\bf[Le87] \sl Lesniewski, A.:} {\it Effective actions for the
Yukawa${}_2$ quantum field theory}, Communications in Mathematical
Physics, {\bf108}, 437- 467, 1987.

\0{\bf[LL63] \sl Lieb E., Lininger, W.:} {\it Exact analysis of an
interacting Bose gas. I. The general solution and the ground state},
Physical Review, {\bf 130}, 1606--1615, 1963, reprinted in [ML66]. See
also {\sl Lieb E.:} {\it Exact analysis of an interacting Bose
gas. I. The excitation spectrum}, Physical Review, {\bf 130},
1616--1624, 1963.

\0{\bf[Lu60] \sl Luttinger, J.M.:} {\it Fermi surface and some simple
equilibrium properties of a system of interacting fermions}, Physical
Review, {\bf119}, 1153--1163, 1960.

\0{\bf[Lu63] \sl Luttinger, J.M.:} {\it An exactly soluble model of a
many fermion system}, Journal of Mathematical Physics, {\bf4},
1154-\-1162, 1963.

\0{\bf[LW60] \sl Luttinger, J.M., Ward, J.:} {\it Ground state energy
of a many fermion system. II}, Physical Review {\bf 118},
1417--1427, 1960.

\0{\bf[LW68] \sl Lieb, E., Wu, F.Y.:} {\it Absence of Mott transition
in an exact solution of the short-range, one-band model in
one-dimension}, Physical Review Letters, {\bf 20}, 1445--1448, 1968.

\0{\bf[LY00] \sl Lieb, E., Yngvason, J.:} {\it The Ground State Energy 
of a Dilute Two-dimensional Bose Gas}, preprint 00-63 in {\tt
mp$\_$arc@ma.utexas.ed}, 2000.

\0{\bf[Ma00] \sl Mastropietro, V.:} private communication.

\0{\bf[Ma64] \sl Mattis, D.:} {\it Band theory of magnetism in metals
in context of exactly soluble model}, Physics, {\bf1}, 183-\-193,
1964.
\1

\0{\bf[Ma91] \sl Mastropietro, V.:} {\it Interacting soluble Fermi
systems in one dimension}, {\it Nuovo Cimento} {\bf B 1}, 304--312,
1994.

\0{\bf[Ma99a] \sl Mastropietro, V.:} {\it Anomalous BCS equation for 
a Luttinger superconductor}, Modern Physics Letters, {\bf 13},
585--597, 1999. And {\it Anomalous superconductivity for coupled
Luttinger liquids}, in print on Reviews in Mathematical Physics,
preprint 1999, in {\tt http://ipparco.roma1. infn.it}.

\0{\bf[Ma99b] \sl Mastropietro, V.:} {\it A renormalization group
computation of the $XYZ$ correlation functions}, Letters in
Mathematical Physics, {\bf47}, 339--352, 1999.

\0{\bf[Ma00] \sl Mastropietro, V.:} private communication.

\0{\bf[MD93] \sl Metzner, W., Di Castro, C.:} {\it Conservation laws
and correlation functions in the Luttinger liquids}, Physical Review
{\bf B 47}, 16107--16123, 1993.

\0{\bf[ML65] \sl Mattis, D., Lieb, E.:} {\it Exact solution of a many
fermion system and its associated boson field}, Journal of
Mathematical Physics, {\bf6}, 304--310?, 1965 (reprinted in: [ML66]).

\0{\bf[ML66] \sl Mattis, D., Lieb, E.:} {\it Mathematical Physics in
one dimension}, Academic Press, New York, 1966.

\0{\bf[Ba82] \sl Baxter, R.:} {\sl Exactly solved models}, Academic Press,
London, 1982.

\0{\bf[Sh92] \sl Shankar, R.:} {\it Renormalization group approach to
interacting fermions}, Review of Modern Physics, {\bf 66}, 129--192,
1994.

\0{\bf[Si74] \sl Simon, B.:} {\it The $P(\f)_2$ euclidean (quantum)
field theory}, Princeton University Press, 1974.

\0{\bf[So79] \sl S\'olyom, J.:} {\it The Fermi gas model of
one-dimensional conductors}, Advances in Phys\-ics {\bf 28}, 201--303,
1979.

\0{\bf[Sp99] \sl Spencer, T.:} {\it A mathematical approach to
universality in two dimensions,} preprint, 1999.

\0{\bf[To50] \sl Tomonaga, S.:} {\it Remarks on Bloch's method of
sound waves applied to many fermion problems}, Progress of Theoretical
Physics, {\bf5}, 349--374, 1950, (reprinted in [ML66]).
}
\*

\FINE
\end